\documentstyle[12pt,epsf]{article}
\makeatletter \ifcase\@ptsize \font\teneufm=eufm10
\font\seveneufm=eufm7 \font\fiveeufm=eufm5
\font\teneusm=eusm10 \font\seveneusm=eusm7
\font\fiveeusm=eusm5 \or \font\teneufm=eufm10 scaled
\magstephalf \font\seveneufm=eufm7 \font\fiveeufm=eufm5
\font\teneusm=eusm10 scaled \magstephalf
\font\seveneusm=eusm7 \font\fiveeusm=eusm5 \or
\font\teneufm=eufm10 scaled \magstep1 \font\seveneufm=eufm7
\font\fiveeufm=eufm5 \font\teneusm=eusm10 scaled \magstep1
\font\seveneusm=eusm7 \font\fiveeusm=eusm5 \fi

\newfam\eufmfam \newfam\eusmfam \textfont\eufmfam=\teneufm
\scriptfont\eufmfam=\seveneufm
\scriptscriptfont\eufmfam=\fiveeufm
\textfont\eusmfam=\teneusm \scriptfont\eusmfam=\seveneusm
\scriptscriptfont\eusmfam=\fiveeusm

\def\frak{\ifmmode\let\next\frak@\else
 \def\next{\errmessage{Use \string\frak\space only in math
 mode}}\fi\next} \def\frak@#1{{\frak@@{#1}}}
 \def\frak@@#1{\fam\eufmfam#1} 
 \def\sh{\ifmmode\let\next\sh@\else
 \def\next{\errmessage{Use \string\sh\space only in math
 mode}}\fi\next} \def\sh@#1{{\sh@@{#1}}}
 \def\sh@@#1{\fam\eusmfam#1}

\ifcase\@ptsize \font\tenmsa=msam10 \font\sevenmsa=msam7
 \font\fivemsa=msam5 \font\tenmsb=msbm10
 \font\sevenmsb=msbm7 \font\fivemsb=msbm5 \or
 \font\tenmsa=msam10 scaled \magstephalf
 \font\sevenmsa=msam7 \font\fivemsa=msam5
 \font\tenmsb=msbm10 scaled \magstephalf
 \font\sevenmsb=msbm7 \font\fivemsb=msbm5 \or
 \font\tenmsa=msam10 scaled \magstep1 \font\sevenmsa=msam7
 \font\fivemsa=msam5 \font\tenmsb=msbm10 scaled \magstep1
 \font\sevenmsb=msbm7 \font\fivemsb=msbm5 \fi

\newfam\msafam \newfam\msbfam \textfont\msafam=\tenmsa
\scriptfont\msafam=\sevenmsa
\scriptscriptfont\msafam=\fivemsa \textfont\msbfam=\tenmsb
\scriptfont\msbfam=\sevenmsb
\scriptscriptfont\msbfam=\fivemsb

\def\Bbb{\ifmmode\let\next\Bbb@\else
 \def\next{\errmessage{Use \string\Bbb\space only in math
 mode}}\fi\next} \def\Bbb@#1{{\Bbb@@{#1}}}
 \def\Bbb@@#1{\fam\msbfam#1} \def\hexnumber@#1{\ifnum#1<10
 \number#1\else \ifnum#1=10 A\else\ifnum#1=11
 B\else\ifnum#1=12 C\else \ifnum#1=13 D\else\ifnum#1=14
 E\else\ifnum#1=15 F\fi\fi\fi\fi\fi\fi\fi}
 \def\msa@{\hexnumber@\msafam} \def\msb@{\hexnumber@\msbfam}
 \mathchardef\square="0\msa@03

\makeatother
 \newcommand{\RR}{{\Bbb R}}
 
\newcommand{\ZZ}{{\Bbb Z}}

\newtheorem{Lemma}{Lemma}[section]
\newtheorem{Proposition}{Proposition}[section]

\def\proof{\par{\it Proof}. \ignorespaces}
\def\endproof{{\ \vbox{\hrule\hbox{%
   \vrule height1.3ex\hskip0.8ex\vrule}\hrule }}\par}
\newenvironment{Proof}{\proof}{\endproof}

\begin{document}

\title{Toda lattices with indefinite metric II:\\
Topology of the iso-spectral manifolds}

\author{Yuji Kodama\thanks{
{\it E-mail }: kodama@math.ohio-state.edu}
\\
\it Department of Mathematics,  Ohio State University\\
\it Columbus, OH 43210, USA \\
and \\
\it Department of Communication Engineering\\
\it Osaka University,
\it Suita, Osaka 565, Japan
\\
\\
 Jian Ye\thanks{
{\it E-mail }: ye@math.ohio-state.edu}
\\
\it Department of Mathematics, Ohio State University\\
\it Columbus, OH 43210, USA \\
}

\maketitle
\begin{abstract}

We consider the iso-spectral
{\it real} manifolds of tridiagonal Hessenberg matrices with
real eigenvalues. The manifolds are described by the iso-spectral
flows of indefinite Toda lattice equations introduced by the authors
[{\it Physica}, {\bf 91D} (1996), 321-339].  These Toda lattices consist
 of $2^{N-1}$ different systems with hamiltonians
$H=\frac{1}{2}\sum_{k=1}^{N}y_k^2 + \sum_{k=1}^{N-1}s_ks_{k+1}
\exp(x_k-x_{k+1})$, where $s_i=\pm 1$. We compactify the manifolds by adding
infinities according to the Toda flows which blow up in finite time
except the case with all $s_is_{i+1}=1$.
The resulting manifolds are shown to be nonorientable for
$N>2$, and the symmetric group is
the semi-direct product of $(\ZZ_2)^{N-1}$ and the permutation group $S_N$.
These properties identify themselves with ``small covers'' introduced by
Davis and Januszkiewicz [{\it Duke Mathematical Journal}, {\bf 62} (1991),
417-451].
As a corollary of our construction, we give a formula on the total
numbers of zeroes for a system of exponential polynomials generated as
Hankel determinants.
\end{abstract}


\section{Introduction}
\renewcommand{\theequation}{1.\arabic{equation}}\setcounter{equation}{0}

The finite non-periodic Toda lattice equation \cite{moser} describes
a hamiltonian system of $N$ particles on a line interacting pairwise with
exponential forces. The hamiltonain $H_T$ of the system is given by
\begin{eqnarray}
\label{ham}
H_T=\frac{1}{2} \sum_{k=1}^{N} y_k^2 + \sum_{k=1}^{N-1} \exp(x_k-x_{k+1}).
\end{eqnarray}
Flaschka \cite{flash} introduced a change of variables
\begin{eqnarray}
\label{a}
a_k=-\frac{y_k}{2},  \quad  k=1, \cdots, N,
\end{eqnarray}
and
\begin{eqnarray}
\label{b0}
b_k=\frac{1}{2} \exp \left(\frac{x_k-x_{k+1}}{2} \right),  \quad k=1,\cdots,
N-1,
\end{eqnarray}
to write the Toda equation in the Lax form:
\begin{eqnarray}
\label{toda}
\frac{d }{dt} L_T = \left[ B_T \ , \ L_T \right] \ ,
\end{eqnarray}
where $L_T$ is an $N \times N$ symmetric ``tridiagonal''
matrix with real entries,
\begin{eqnarray}
\label{l0}
L_T = \left(
\begin{array}{lllll}
a_1 & b_1 & 0 & \cdots & 0 \\
b_1 & a_2 & b_2 & \cdots & 0 \\
\vdots & \ddots & \ddots & \ddots & \vdots \\
0 & \cdots & \ddots & a_{N-1} & b_{N-1} \\
0 & \cdots & \cdots & b_{N-1} & a_N \\
\end{array}
\right)
\end{eqnarray}
and $B_T$ is the
skew symmetric matrix defined by
\begin{eqnarray}
\label{skew0}
B_T = \prod_{a} L_T := \left( L_T \right)_{>0} -
\left( L_T \right)_{<0} \ .
\end{eqnarray}
Here $\left( L_T \right)_{>0 \ (<0)}$ denotes the strictly upper (lower)
triangular part of $L_T$.
As an immediate consequence of the Lax form (\ref{toda}), the eigenvalues
of $L_T(t)$ are time-invariant. Moreover, as $t \rightarrow \infty$,
$L_T(t)$  converges to a diagonal matrix with eigenvalues of
$L_T$ in the
decreasing order \cite{moser}, which is referred as ``the sorting peoperty''.

\medskip

Tomei \cite{tomei} used the Toda flow (\ref{toda}) to study the iso-spectral
manifolds of symmetric tridiagonal matrices, i.e., in the form of
(\ref{l0}). He showed several interesting properties of the manifolds:
The manifolds are orientable, with Euclidean $\RR^{N-1}$ as the
universal covering,
and the symmetry group is $(\ZZ_2)^{N-1} \times S_N$, i.e., the the direct
product of $(\ZZ_2)^{N-1}$ and the permutaion group $S_N$. Davis \cite{davis}
generalized Tomei's manifolds using reflection groups acting on some
simple polytope.

\medskip

In \cite{ky}, the authors considered equation (\ref{toda}) replacing
$L_T$ with
$\tilde L=LS$ and $B_T$ with $\tilde B=\prod_{a}\tilde L$,
where $L$ is full symmetric and $S$ is a diagonal matrix,
$S=diag(s_1,\cdots,s_N)$ with $s_i=\pm 1$. An explicit solution formula
is given by extending the ``orthonormalization'' method introduced in
\cite{km}. In particular, for the case of tridiagonal $\tilde L(=L_TS)$, in
analogue to the Toda equation (\ref{toda}), the corresponding
hamiltonian $H$ is given by
\begin{eqnarray}
\label{ham1}
 H=\frac{1}{2}\sum_{k=1}^{N}y_k^2 +
\sum_{k=1}^{N-1}s_ks_{k+1}\exp(x_k-x_{k+1}).
\end{eqnarray}
In this paper, we consider the integral manifolds of the hamiltonian
systems given by (\ref{ham1}).
With different signs of $s_i$'s, there are a total of $2^{N-1}$ different
type of Toda lattices. We shall refer them except the one with all
positive or negative $s_i$'s as ``indefinite Toda lattices''.
Note from (\ref{ham1}) that $ H$
includes some attractive forces, thereby is not positive definite.
One then expects a `` blowing up'' in solutions.
Indeed, it was shown in \cite{ky} that generically there are two types
of solutions, having either the sorting property or blowing up to infinity
in finite time. The structure of $\tau$-functions describing
the solutions was also studied,
and they are not
positive definite as for the original Toda equation.

\medskip

Indefinite Toda lattices with the hamiltonian (\ref{ham1}) arise in the
symmetry reduction of
the so-called {\it Wess-Zumino-Novikov-Witten} model \cite{ft},
which is one of the most important models in quantum
field
theory and string theory.  It was then shown
\cite{ft} that
the reduced system contains all the indefinite Toda lattices. Even though
each of these Toda lattices is singular in the sense of blow-up
in the solution, these are expected
to be regularized by gluing them together through ``infinities'' in the
full reduced system. The main purpose of the present
 paper is to give a concrete description
of such regularization. Namely we study a compactification
of the integral manifolds of these indefinite Toda lattices.

\medskip

Our method is based on the explicit solution given in \cite{ky}, which
is different from
the approach based on the Bruhat decomposition. The Bruhat
decomposition of the special linear group $G$ is
\begin{eqnarray}
\label{bruhat}
G=\bigcup_{w\in W}N_-wB_+\ ,
\end{eqnarray}
where $N_-$ is the unipotent subgroup
 of lower triangular matrices with 1's on
the diagonal, $B_+$ is the Borel subgroup of upper triangular matrices,
and $W$ is the Weyl group of $G$.
The cell of maximal dimension is $N_-B_+$ (i.e. $w=id$)
which corresponds to the
LU factorization. It is shown  that the blow-up
of the indefinite Toda flow with $\tilde L \in G$ at
$t_0$ corresponds to  the intersection,
\begin{eqnarray}
\label{el0}
e^{t_0\tilde L(0)} \in N_-wB_+, \quad {\mbox {where}} \ \ w \neq id \ .
\end{eqnarray}
The codimension of $N_-wB_+$
(which is given by the length of $w$) then gives the degeneracy of the blow-up.
The complex version of the decomposition has been studied by Flaschka and
Haine \cite{fh1,fh2} where they use ``Painl\'eve analysis'' to study the
singularities and to compactify the flag manifold $G/B_{+}$.

\medskip

The paper is organized as follows. In Section 2, we give a summary of the
results obtained in \cite{ky} and reference therein for a basic
background of the present paper. In Section 3, we provide a detail
study of the solution behaviors. In particular, we give the total number
of blow-ups for generic orbits, which
extends the well known result of the total positivity
of $\tau$-functions studied in \cite{karlin}.
In Section 4, we give an explicit construction
of gluing indefinite Toda lattices by adding infinities, i.e.,
compactification of the iso-spectral manifold. We start
with the
cases of $N=2$ and $3$ as examples, then proceed to the  general case.
In Section 5, we give a brief account of CW decomposition of the
compacified iso-spectral manifold and
show its nonorientability.

\section{Toda lattice with indefinite metric}
\renewcommand{\theequation}{2.\arabic{equation}}\setcounter{equation}{0}

For the hamiltonian (\ref{ham1}), we introduce a change of variables
due to Flaschka \cite{flash}
\begin{eqnarray}
\label{as}
s_ka_k&=&-\frac{y_k}{2},  \quad \quad  k=1, \cdots, N, \\
\label{bs}
b_k&=&\frac{1}{2}\exp\left({x_k-x_{k+1} \over 2}\right),
  \quad  k=1, \cdots, N-1 \ ,
\end{eqnarray}
and $t \rightarrow t/2$ to write hamilton's equations in the
following form with $b_0=b_N=0$,
\begin{eqnarray}
\label{exa}
\frac{d a_k}{d t}&=&s_{k+1}b_k^2-s_{k-1}b_{k-1}^2,\\
\label{exb}
\frac{d b_k}{d t}&=&{1 \over 2}b_k(s_{k+1}a_{k+1}-s_ka_k).
\end{eqnarray}
This system can be expressed in the Lax form,
\begin{eqnarray}
\label{todals}
\frac{d}{d t} L
= \left[ B\ , L \right] \ ,
\end{eqnarray}
where $L$ is an $N \times N$ tridiagonal matrix with real entries,
\begin{eqnarray}
\label{l}
L = \left(
\begin{array}{lllll}
s_1a_1 & s_2b_1 & 0 & \cdots & 0 \\
s_1b_1 & s_2a_2 & s_3b_2 & \cdots & 0 \\
\vdots & \ddots & \ddots & \ddots & \vdots \\
0 & \cdots & \ddots & s_{N-1}a_{N-1} & s_Nb_{N-1} \\
0 & \cdots & \cdots & s_{N-1}b_{N-1} & s_Na_N \\
\end{array}
\right)
\end{eqnarray}
and
$B$ is the projection of $L$ given by
\begin{eqnarray}
\label{skew}
B = \frac{1}{2}\prod_{a} L := \frac{1}{2}\left[\left( L \right)_{>0} -
\left( L \right)_{<0}\right] \ .
\end{eqnarray}
Note from (\ref{l}), $L=L_TS$ where $L_T$ is a symmetric tridiagonal
matrix (\ref{l0}) and $S$ is a diagonal matrix $S=diag(s_1,\cdots,s_N)$.
In \cite{ky}, more general situation where $L_T$ is a ``full''
symmetric matrix is considered. An explicit formula for the solution is
given based on the inverse scattering method. Here we give a summary of
the results obtained in \cite{ky}.

\medskip

The inverse scattering scheme for (\ref{todals})
consists of two linear equations,
\begin{eqnarray}
\label{comp1}
L \Phi &=& \Phi \Lambda  \ , \\
\label{comp2}
\frac{ d }{d t} \Phi &=& B \Phi \ ,
\end{eqnarray}
where $\Phi$ is the eigenmatrix of $L$, and $\Lambda$ is $diag
\left( \lambda_1, \cdots, \lambda_N \right)$.
It is shown that $\Phi$ can be normalized to satisfy
\begin{eqnarray}
\label{PhiS}
\Phi S^{-1} \Phi^T \ = S^{-1}, \ \ \Phi^T S \Phi \ = S.
\end{eqnarray}
In particular, if $S=I$, the identity matrix, then (\ref{PhiS}) implies
that
$L$ can be diagonalized by an orthogonal matrix $O(N)$, and if
$S=diag(1,\cdots,1,-1,\cdots,-1)$ then the diagonalization
is obtained by
a pseudo-orthogonal matrix $O(p,q)$ with $p+q=N$.
It should be noted that with the normalization (\ref{PhiS}),
the eigenmatrix $\Phi$ becomes complex
in general, even in the case that all the eigenvalues are real. This is simply
due to a case where the sign of $\sum_{k=1}^Ns_k\phi_k^2(\lambda_i)$
in $\Phi^TS\Phi$ differs
from that of $s_i$.

\medskip

The eigenmatrix $\Phi$ consists of the eigenvectors of $L$,
$L\phi=\lambda \phi$,
with $\phi(\lambda_{k}) \equiv ( \phi_{1}(\lambda_{k}),\\\cdots $,
$\phi_{N}(\lambda_{k} ))^{T} $ for $k = 1, 2, \cdots , N$,
\begin{eqnarray}
\label{orth}
\Phi \equiv \left[ \phi(\lambda_{1}), \ \cdots \ , \ \phi(\lambda_{N})
\right] \ = \ \left[ \phi_{i}(\lambda_{j}) \right]_{1 \le i , j \le N} \ .
\end{eqnarray}
Then (\ref{PhiS}) give the ``orthogonality'' relations
\begin{eqnarray}
\label{ortho}
\sum_{k=1}^{N} s_k^{-1}\phi_{i}(\lambda_{k}) \phi_{j}(\lambda_{k}) &=&
\delta_{ij} s_i^{-1} \ , \\
\sum_{k=1}^{N} s_k\phi_{k}(\lambda_{i}) \phi_{k}(\lambda_{j}) &=&
\delta_{ij}s_i \ .
\end{eqnarray}
With (\ref{ortho}), we now define an inner product $<\cdot,\cdot>$ for
two functions $f$ and $g$ of $\lambda$ as
\begin{eqnarray}
\label{product}
<f, g> := \sum_{k=1}^Ns_k^{-1}f(\lambda_k)g(\lambda_k),
\end{eqnarray}
which we write as $<fg>$ in the sequel.
The metric in the inner product is given by
\begin{eqnarray}
\label{metric}
d \alpha (\lambda) = \sum_{k=1}^{N} s_k^{-1}\delta(\lambda - \lambda_{k}) d
\lambda \ ,
\end{eqnarray}
which leads to an indefinite metric due to a choice of negative entries
$s_k$ in $S$.
The entries of $L$ are then expressed by
\begin{eqnarray}
\label{back}
 a_{ij} := \left( L \right)_{ij} =
 s_j < \lambda \phi_{i} \phi_{j} > \ .
\end{eqnarray}

\medskip

The time evolution of $\Phi(t)$ can be obtained by the
orthonormalization procedure of Szeg\"o \cite{szego} with respect to
the metric (\ref{product}). This also generalizes the orthonormalization
method introduced by Kodama and McLaughlin \cite{km}. An explicit form
 of $\Phi(t)$ is then given by
\begin{eqnarray}
\label{evcs}
\phi_{i}(\lambda , t) = \frac{e^{\lambda t}}
{\sqrt{D_{i}( t)
D_{i-1}(t)}} \left|
\begin{array}{ccccc}
s_1c_{11} & s_2c_{12} & \ldots & s_{i-1}c_{1,i-1} & \phi_{1}^{0}(\lambda) \\
\vdots & \vdots & \ddots & \vdots & \vdots \\
s_1c_{i1} & s_2c_{i2} &\ldots & s_{i-1}c_{i,i-1} & \phi_{i}^{0}(\lambda) \\
\end{array}
\right|,
\end{eqnarray}
where $\phi^0_i(\lambda):=\phi_i(\lambda,0)$,
 $c_{ij}(t) = < \phi_{i}^{0} \phi_{j}^{0} e^{\lambda t}>$, and
$D_{k}(t)$ is the determinant of the
$k \times k$ matrix with entries $s_ic_{ij}(t)$, i.e.,
\begin{eqnarray}
\label{DDD}
D_{k}( t) = \left| \Big( s_ic_{ij}(t) \Big)_{1 \le i,j \le k}
\right| \ .
\end{eqnarray}
The determinants $D_k(t)$ are normalized as $D_k(0)=1$ for all $k$.
With the formula (\ref{evcs}), we now have the solution (\ref{back})
of the inverse scattering problem (\ref{comp1}) and (\ref{comp2}).

\medskip

It is immediate from the explicit formula (\ref{evcs}) that
\begin{Proposition}.
\label{zero}
Suppose $D_i(t_0)=0$ for some $t_0$ and some $i$, then
$ L(t)$ blows up to infinity at $t_0$.
\end{Proposition}
In the tridiagonal case, the determinants $D_i(t)$'s can be written as
the so-called $\tau$-functions.
Then the solutions  $a_i$'s and $b_i(t)$'s are expressed in the form,
\begin{eqnarray}
\label{atau}
 s_ia_i &=& \frac{d}{d t} {\log \frac{\tau_i}{\tau_{i-1}}}, \\
\label{btau}
s_is_{i+1} b_i^2&=&\frac{\tau_{i+1} \tau_{i-1}}{\tau_i^2}.
\end{eqnarray}
The indefinite Toda lattice equations (\ref{exa}) and (\ref{exb}) are written
in the {\it bilinear} form,
\begin{eqnarray}
\label{etau}
\frac{d^2}{d t^2}\log \tau_i=\frac{\tau_{i+1}\tau_{i-1}}{\tau_i^2}.
\end{eqnarray}
These $\tau$-functions $\tau_i$ have a simple structure, that is,
a Hankel determinant given by \cite{by,fay,nk}
\begin{eqnarray}
\label{tau}
\tau_i = \left|
\begin{array}{lllll}
\tau_1 & \tau_1{'} & \tau_1{''} & \ldots & \tau_1^{(i-1)}\\
\tau_1' & \tau_1{''} & \tau_1^{(3)} & \ldots & \tau_1^{(i)} \\
\vdots  & \vdots & \ddots & \vdots & \vdots  \\
\tau_1^{(i-1)} & \tau_1^{(i)} & \ldots & \ldots & \tau_1^{(2i-2)}\\
\end{array}
\right|,
\end{eqnarray}
where $\tau_1$ is given by $\tau_1=c_{11}:=<(\phi_1^0)^2e^{\lambda t}>=
s^{-1}_1D_1$, and $\tau^{(i)}_1=d^i \tau_1 /dt^i$.
The relation between $\tau_i$ and $D_i$ is given by
\begin{eqnarray}
\label{taud}
\tau_i=\frac{1}{s_1^i}\left[\prod_{k=1}^{i-1}(s_ks_{k+1}(b_k^0)^2)^{i-k}
\right] D_i\ ,
\end{eqnarray}
where $b_k^0=b_k(0)$. (The formula (6.5) in \cite{ky} should read
as (\ref{taud}).)

\medskip

{\bf Remark 1}. From (\ref{taud}), $\tau_i$ becomes $0$ if $b_k^0=0$
for some $k<i$, thereby
(\ref{btau}) is no longer valid. This is because the system can be reduced
to two or more independent subsytems.

\medskip

{\bf Remark 2}. The system (\ref{todals}) can be also written in the Lax
form with
a tridiagonal Hessenberg matrix with $\alpha_i:=s_ia_i$ and $\beta_i:=s_is_{i+1}
b_i^2$,
\begin{eqnarray}
\label{lh}
L_H = \left(
\begin{array}{lllll}
\alpha_1 & 1  & 0 & \cdots & 0 \\
\beta_1 & \alpha_2 & 1 & \cdots & 0 \\
\vdots & \ddots & \ddots & \ddots & \vdots \\
0 & \cdots & \ddots & \alpha_{N-1} & 1 \\
0 & \cdots & \cdots & \beta_{N-1} & \alpha_N \\
\end{array}
\right).
\end{eqnarray}
The matrix $L_H$ is similar to $L$ in (\ref{l}), $L_H=HLH^{-1}$, with the
diagonal matrix $H=diag(1, s_2b_1,s_2s_3b_1b_2,\cdots,\prod_{i=1}^{N-1}
s_{i+1}b_i)$ \cite{ky-cmp}.
Then (\ref{todals}) becomes
\begin{eqnarray}
\label{todah}
\frac{d }{d t} L_H = \left[ B_H \ , \ L_H \right] \ ,
\end{eqnarray}
where $B_H=(L_H)_{<0}$, i.e., the lower triangular part of $L_H$.
Note from (\ref{btau}), if $\tau_{i-1}$ or $\tau_{i+1}$ changes sign,
then $b_i^2$ becomes negative. In this paper we use the variables $\alpha_i$ and
$\beta_i$, and consider the {\it real} form of the iso-spectral manifold
of the system of equations (\ref{exa}) and (\ref{exb}).

\medskip

{\bf Remark 3.} With the Hessenberg matrix form of (\ref{todah}),
the Toda equation with the general tridiagonal matrix can be solved
in the similar way \cite{ky-cmp}.

\medskip

\section{Behaviors of solutions}
\renewcommand{\theequation}{3.\arabic{equation}}\setcounter{equation}{0}

We use the $\tau$-functions (\ref{tau}) to study the behaviors of solutions
 for all $t \in {\Bbb R}$. For $t>0$,
a neccessary and sufficient condition for solutions being nonsingular is
obtained in \cite{GS}.
First we have the following forms of $\tau$-functions as sums of
exponential functions:
\begin{Proposition}
\label{expansion}
Write $\tau_1=\sum_{k=1}^{N}\rho_ie^{\lambda_i t}$ where $\rho_i =s_i[\phi_1^0
(\lambda_i)]^2\neq 0$. Then
$\tau_i$ for $i = 1, 2, \cdots, N$ in (\ref{tau}) can be expressed as
\begin{eqnarray}
\label{expand}
\tau_i(t)= \sum_{J_N=(j_1,\cdots,j_i)}
\rho_{j_1}\rho_{j_2}\cdots \rho_{j_i}
 \left|
\begin{array}{ccc}
1 & \cdots & 1 \\
\vdots & \ddots & \vdots \\
\lambda_{j_1}^{i-1} & \cdots & \lambda_{j_{i}}^{i-1} \\
\end{array}
\right|^2 \exp (\sum_{k=1}^i\lambda_{{j_k}}t),
\end{eqnarray}
where $J_N$ represents all possible combinations for
$1\le j_1<\cdots<j_i\le N$.
In particular $\tau_N(t)=\rho_1\cdots \rho_N
\prod_{i<j}(\lambda_i-\lambda_j)^2 \exp (\sum_{k=1}^N\lambda_kt)$.
\end{Proposition}

If $\beta_i^0:=s_is_{i+1}(b_i^0)^2>0$ for
$i=1,\cdots,N$, then  all $\rho_i>0$, and
 from (\ref{expand}) we see that $\tau_i(t)$'s are all
positive definite.
Conversely, if $\tau_i(t)$ are all
positive definite, then $\beta_i(t)>0$ for all $t$ by
(\ref{btau}).
On the other hand, if $\rho_k<0$ for some $k$, then the existence of zeroes
for $\tau_i(t)$'s is guaranteed by Karlin \cite{karlin}.
We summarize these facts as:

\begin{Proposition}
\label{ezero}
Write $\tau_1=\sum_{k=1}^{N}\rho_ie^{\lambda_i t}$, $\rho_k\neq 0$ for
$k=1,\cdots,N$.
Then $\tau_k(t)$ are all sign definite, i.e. have no zeroes,
for $t \in {\Bbb R}$ if and only if either
$\rho_k>0$ or $\rho_k<0$ for all $k$.
\end{Proposition}

The above proposition implies the Toda flow is neccessarily singular
for $t\in {\Bbb R}$ if
$\beta_k^0 < 0$ for some $k$.
Moreover, we can give the precise number of blow-ups for generic orbits
based on the
construction in the next section.

\medskip

\noindent
{\bf Definition} A zero $t_0$ of $\tau_i(t)$'s is said to be {\it
nondegenerate} if $\tau_k(t_0)=0$ for some $k$ and $\tau_j(t_0) \neq 0$
for $j\neq k$.

\medskip

Note that (\ref{etau}) can be written as
\begin{eqnarray}
\label{jacobi}
\tau_i\tau_i''-(\tau_i')^2=\tau_{i+1}\tau_{i-1} .
\end{eqnarray}
If $\tau_i$ has multiple zeroes at $t_0$, then $\tau_{i+1}$ or $\tau_{i-1}$
must be zero at $t_0$. So a zero of multiplicity more than 1 is
necessarily degenerate. From the
above definition, a zero is nondegenerate if only one $\tau$-function
reaches zero. Note that nondegenerate zeroes are generic.  Then we have:

\begin{Proposition}
\label{nzero}
Write $\tau_1=\sum_{k=1}^{N}\rho_ie^{\lambda_i t}$.
Suppose that all zeroes of $\tau$-functions
are nondegenerate.
Denote by $m$ the number of negative $\rho_i$'s, and by $n_i$
the number of zeroes of $\tau_i(t)$. Then the total number of the zeroes
depends only on $m$ and is given by
\begin{eqnarray}
\label{zeroes}
\sum_{i=1}^{N-1}n_i=m(N-m)\ .
\end{eqnarray}
\end{Proposition}
Note here that the number of negative $\rho_i$'s takes at most
$[N/2]$, the integer part of $N/2$, because of the symmetry of
$\alpha_i$ and $\beta_i$ under the change $\tau_1 \to -\tau_1$.
Thus the total number of zeros is a topological quantity of the
$\tau$-functions, and this Proposition extends a result of Karlin
\cite{karlin}.
The proof will be given in the next section.

Though there are generic orbits that blow up to infinity in finite time,
asymptotically, we have:

\begin{Proposition}
\label{sorting}
Suppose that $\beta_i^0 \neq 0$ for $i=1, \cdots, N-1$, and
$\lambda_1> \cdots >\lambda_N$. Then  $L(t) \rightarrow
diag(\lambda_1,\cdots \lambda_N)$ as $t \rightarrow
\infty$, and
 $L(t) \rightarrow diag(\lambda_N,\cdots \lambda_1)$ as $t \rightarrow
-\infty$.
\end{Proposition}
\begin{Proof}
Since $\beta_i^0 \neq 0$ for all $i=1$, $\tau_1(t)$ is
given by $\sum_{k=1}^{N}\rho_ie^{\lambda_it}$ with $\rho_i \neq 0$.
We calculate directly asymptotic behavior of solution using
(\ref{atau}) and (\ref{btau}).
For large $t$, we have
\begin{eqnarray}
\nonumber
\alpha_i(t)&=&\frac{d}{dt}\log{\frac{\tau_i}{\tau_{i-1}}} \\
\nonumber
&\rightarrow& \frac{d}{dt}\log{\frac{\rho_1\cdots \rho_i \prod_{1<k<l<i}
(\lambda_k-\lambda_l)^2e^{ \sum_{k=1}^i\lambda_{{k}}t}}
{\rho_1\cdots \rho_{i-1} \prod_{1<k<l<i-1}
(\lambda_k-\lambda_l)^2e^{ \sum_{k=1}^{i-1}\lambda_k t}}} =\lambda_i, \\
\nonumber
\beta_i(t)&=&\frac{\tau_{i+1}\tau_{i-1}}{\tau_i^2} \\
\nonumber
&\rightarrow& \gamma e^{(\lambda_{i+1}-\lambda_i)t}
\rightarrow 0\ ,
\end{eqnarray}
where $\gamma$ is a nonzero constant.
Similarly, one can show  $L(t) \rightarrow diag(\lambda_N,\cdots \lambda_1)$ as
$t \rightarrow -\infty$.
\end{Proof}

\medskip

\section{Topology of indefinite Toda lattices}
\renewcommand{\theequation}{4.\arabic{equation}}\setcounter{equation}{0}

In this section, we give an explicit construction of
a compactification (or regularization) of the integral manifolds
of the indefinite Toda lattices.
For this purpose, we use $L_H$ given in (\ref{lh})
with all eigenvalues being real and distinct as describing
the underlying manifolds. We
assume that the eigenvalues of $L_H$ are ordered as $\lambda_1>
\cdots > \lambda_N$. We associate $L_H$ with an $S$ matrix through
$sgn(\beta_i)=s_is_{i+1}$ as in (\ref{lh}). With different signs
of $\beta_i$, there are a total of $2^{N-1}$ different Toda lattices.
As we show here, the compactification is then given by
gluing these different Toda lattices.

\medskip

First we study the simplest case with $N=2$ which is also the most important
case as we shall see later. Then we give a detail discussion on
the case $N=3$, and extend those results to the general case.
\medskip

Let  $L_H$  be a $2\times 2$ Hessenberg
matrix given by
\begin{eqnarray}
\label{lh22}
L_H =\left(
\begin{array}{cc}
\alpha_1 & 1 \\
\beta_1 & \alpha_2 \\
\end{array}
\right) = HLH^{-1},
\end{eqnarray}
where $L=L_TS$ is given by (\ref{l}) and $H=diag(1,s_2b_1) $ (see Remark 2).
We assume that the eigenvalues of $L_H$ are all real and
 $\lambda_1 > \lambda_2$. The characteristic
polynomial for $L_H$ leads to
\begin{eqnarray}
\nonumber
\alpha_1 + \alpha_2 &=& \lambda_1+\lambda_2 , \\
\label{ab}
\alpha_1\alpha_2-\beta_1&=&\lambda_1\lambda_2 .
\end{eqnarray}
Eliminating $\alpha_2$ in (\ref{ab}), we have
\begin{eqnarray}
\label{b1}
\beta_1=-(\alpha_1-\lambda_1)(\alpha_1-\lambda_2)\ .
\end{eqnarray}
The plot of (\ref{b1}) is shown in Fig. 1.
There are two critical points ($b_1=0$ or $\beta_1=0$) corresponding to
$L=diag(\lambda_1, \lambda_2)$ and $diag(\lambda_2, \lambda_1)$.
Let us now give a detail analysis of the orbits of indefinite Toda
lattices associated with $L_H$ in (\ref{lh22}):

\medskip

(i) The case $S=diag(1,1)=I$:  Then the initial data $L^0=L(0)$
is symmetric and diagonalized by an orthogonal matrix
$\Phi^0=\Phi(0)$, i.e. $\Lambda=(\Phi^0)^{-1}L^0\Phi^0$,
\begin{eqnarray}
\label{phi01}
\Phi^0=\left(
\begin{array}{cc}
\cos\theta_0 & \sin\theta_0 \\
-\sin\theta_0 & \cos\theta_0 \\
\end{array}
\right)\ ,
\end{eqnarray}
where $\theta_0$ is determined by the initial data.  Note that the solution
of the Toda equation depends on one parameter $\theta_0$ with fixed $\lambda_1$
and $\lambda_2$. From (\ref{DDD}) and (\ref{evcs}), we have
\begin{eqnarray}
\label{d11}
D_1(t)=\cos^2\theta_0 \ e^{\lambda_1 t}+\sin^2\theta_0 \ e^{\lambda_2 t}\ ,
\end{eqnarray}
and
\begin{eqnarray}
\label{phit1}
\Phi(t)=\frac{1}{\sqrt{D_1(t)}}
\left(
\begin{array}{cc}
 \cos\theta_0 \ e^{{\lambda_1 \over 2} t} &
 \sin\theta_0 \ e^{{\lambda_2 \over 2} t} \\
-\sin\theta_0 \ e^{{\lambda_2 \over 2} t}
&  \cos\theta_0 \ e^{{\lambda_1 \over 2} t} \\
\end{array}
\right).
\end{eqnarray}
We see from (\ref{d11}) that $D_1(t)$ is positive definite, that is,
the solution is regular (no blow up) and asymptotically
\begin{eqnarray*}
& & \Phi (t) \rightarrow sgn(\cos\theta_0)
\left(
\begin{array}{cc}
1 & 0 \\
0 & 1 \\
\end{array}
\right), \quad {\mbox {as}} \  t \rightarrow \infty, \\
& & \Phi (t) \rightarrow sgn(\sin\theta_0)
\left(
\begin{array}{cc}
0 & 1 \\
-1 & 0 \\
\end{array}
\right), \quad {\mbox {as}}  \ t \rightarrow -\infty.
\end{eqnarray*}
Thus $L(t)$ tends to $diag(\lambda_1,\lambda_2)$ as $t \rightarrow \infty$
and $diag(\lambda_2,\lambda_1)$ as $t \rightarrow -\infty$.
The orbit corresponds to a closed component with both critical points
given as the curve in $\beta_1\ge 0$.

\medskip

(ii) The case $S=diag(1,-1)$: The eigenmatrix $\Phi$ for $L$ is now
in $O(1,1)$. There are two disconnected
components in $\beta_1 \le 0$, and each one connects to either
$L=diag(\lambda_1,\lambda_2)$
or $L=diag(\lambda_2,\lambda_1)$.  We first consider the component connected
to the vertex $L=diag(\lambda_1,\lambda_2)$.

Taking the initial data with $\alpha^0_1>\alpha^0_2$. The
initial eigenmatrix $\Phi^0$ is given by
\begin{eqnarray}
\label{phi02}
\Phi^0=\left(
\begin{array}{cc}
\cosh\mu_0 & \sinh\mu_0 \\
\sinh\mu_0 & \cosh\mu_0 \\
\end{array}
\right)\ ,
\end{eqnarray}
where $\mu_0$ determined by the initial data.  From
(\ref{DDD}) and (\ref{evcs}), we obtain
\begin{eqnarray}
\label{d12}
D_1(t)=\cosh^2\mu_0 \ e^{\lambda_1 t}- \sinh^2\mu_0 \ e^{\lambda_2 t},
\end{eqnarray}
and
\begin{eqnarray}
\label{phit2}
\Phi(t)=\frac{1}{\sqrt{D_1(t)}}
\left(
\begin{array}{cc}
 \cosh\mu_0 \ e^{{\lambda_1 \over 2} t} &
 \sinh\mu_0 \ e^{{\lambda_2 \over 2} t} \\
\sinh\mu_0 \ e^{{\lambda_2 \over 2} t}
&  \cosh\mu_0 \ e^{{\lambda_1 \over 2} t} \\
\end{array}
\right).
\end{eqnarray}
 From (\ref{d12}), $D_1(t)$ has a unique zero at
\begin{equation}
\label{tzero}
t_0=\frac{2}{\lambda_1-\lambda_2}\ln(\tanh\mu_0) < 0.
\end{equation}
For $t>t_0$, the eigenmatrix $\Phi(t)$ is real and $\Phi(t) \to I$ as
$t \to \infty$, while for $t<t_0$  $\Phi(t)$ becomes pure imaginary and
$\Phi(t) \to i \ sgn(\sinh \mu_0) \left(\begin{array}{cc}
0 & 1 \\
1 & 0 \\
\end{array}\right) $ which implies $L(t) \to diag(\lambda_2,\lambda_1)$. Thus
the solutions for $t>t_0$ and $t<t_0$ expresses components connecting
to the vertices $L=diag(\lambda_1,\lambda_2)$ and $diag(\lambda_2,\lambda_1)$
respectively.  We then connect these orbits at infinity, that is, we
{\it compactify} the integral manifold of the flow as shown in Fig.1.
The resulting manifold
is isomorphic to circle $S^1$.  This connection of orbits can be viewed
as a gluing of two different indefinite Toda systems.
  To see this more precisely, we
first note that the solution $L_H(t)$ are determined by the $\tau$-function
$\tau_1(t)$,
\begin{equation}
\label{tau1}
\tau_1(t) =s_1[\phi^0_1(\lambda_1)]^2 e^{\lambda_1 t}
+s_2[\phi^0_1(\lambda_2)]^2 e^{\lambda_2 t} .
\end{equation}
Then the change of $\Phi(t)$ from real to imaginary implies the change
of signs in  $\rho_1:=[\phi^0_1(\lambda_1)]^2$ and
$\rho_2:=[\phi^0_1(\lambda_2)]^2$.  This is then equivalent to the change of
signs in $s_1$ and $s_2$, or the exchange $s_1 \leftrightarrow s_2$,
with fixed signs of $\rho_i$'s.  As we will see below,
this can be naturally generalized for the $N \times N$ case,
that is, in each blow-up corresponding to $\tau_i=0$ we glue two systems
with the $S$-matrices $diag(s_1,\cdots,s_i,s_{i+1},\cdots,s_N)$ and
$diag(s_1,\cdots,s_{i+1},s_i,\cdots,s_N)$.


\begin{figure}
\epsfysize=7cm
\centerline{\epsffile{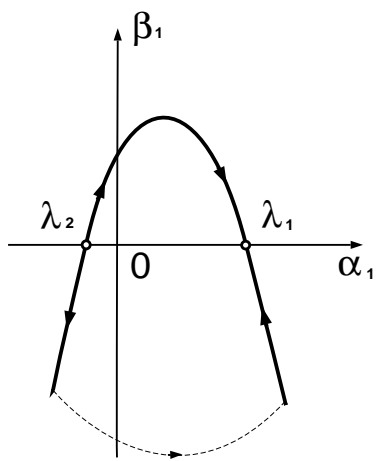}}
\caption{The {\it N=2} indefinite Toda flow.}
\end{figure}

Now let us consider the case  $N=3$ which contains
main ideas for the
general case. For $N=3$, there are six critical points with $\beta_1=
\beta_2=0$ which give the vertices ($0-cells$) of the integral manifold.
Around each vertex, there are four components assigned by the signs
of $\beta_1$ and $\beta_2$, i.e. $++$, $+-$, $-+$, $--$.
 The component ++ is shown to be a hexagon. The edges ($1-cells$) of the
 hexagon correspond to the Toda flows in the
form of either $\beta_1=0$ or $\beta_2=0$. Each vertex has four edges,
two of which are non-compact corresponding to blow-up solutions.
The picture of the underlying manifold is
shown in Fig. 2.


\begin{figure}
\epsfysize=7cm
\centerline{\epsffile{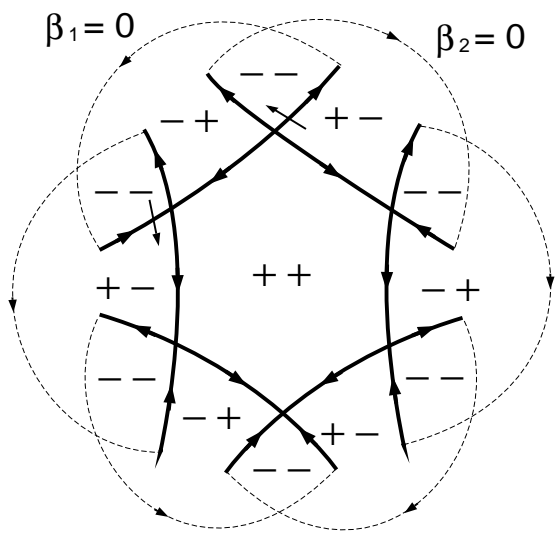}}
\caption{The {\it N=3} indefinite Toda flow.}
\end{figure}

We then compactify the manifold as follows: We first glue the edges as
in the case $N=2$. Then for the flows around an outgoing edge,
we take a small segment transversal to the
flows at some point on the the edge, and determine how the flows transport this
segment after the blow-up. In Fig.2, the gluing pattern around one edge,
$\beta_1=0$, is
indicated by arrows.
Repeating this process for every outgoing edges, we obtain the compactified
manifold.
It is immediate from the gluing pattern that  the manifold
is not orientable.
Its Euler characteristic is calculated as $3-6+1 =-2$ (see below for
a detail).
Since every compact surfaces are completely classified by orietability and
the Euler characteristic, the manifold is topologically isomorphic to
 a connected sum of two Klein bottles.

We can also describe the gluing patterns from the viewpoint of symmetry groups
acting on polytopes \cite{davis,dj} which provides a useful
information for the study of the general
case. We first mark the four components with $++, +-, -+$ and $--$
around the top vertex $V[3,2,1]:=
\left(
\begin{array}{ccc}
\lambda_3 & 1 & 0 \\
0 & \lambda_2 & 1 \\
0 & 0 & \lambda_1 \\
\end{array}
\right)
$ as $A, B, C$ and $D$, respectively.
Then following the flows starting from these components, we mark components
around other vertices. For example, the flows
starting from $B$ blow up and continue to either the component $--$ of
the vertex $V[3,1,2]:=\left(
\begin{array}{ccc}
\lambda_3 & 1 & 0 \\
0 & \lambda_1 & 1 \\
0 & 0 & \lambda_2 \\
\end{array}
\right)
$ or $--$ of $V[2,1,3]:=\left(
\begin{array}{ccc}
\lambda_2 & 1 & 0 \\
0 & \lambda_1 & 1 \\
0 & 0 & \lambda_3 \\
\end{array}
\right)
$, then they blow up again and end in the component $-+$ of the bottom
vertex $V[1,2,3]:=\left(
\begin{array}{ccc}
\lambda_1 & 1 & 0 \\
0 & \lambda_2 & 1 \\
0 & 0 & \lambda_3 \\
\end{array}
\right)
$. We then mark all of these components on the pass as $B$. This gives the
way of gluing the components around each vertex,
and the glued components also form a hexagon. We then have a total of four
hexagons. The gluing pattern and the marking for all components
are shown in Fig.3.  For example, a dotted line segment between a circle
and a square is glued with the other segment which has the same vertices
and connects with the same edge $\beta_k=0$.  So there are 3 vertices,
marked by circle, square and triangle, and 6 edges in the compactified
manifold.


\begin{figure}
\epsfysize=7cm
\centerline{\epsffile{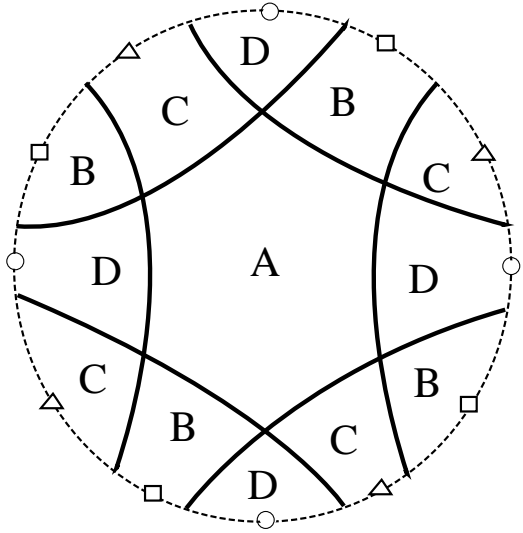}}
\caption{The gluing pattern for the {\it N=3} indefinite Toda manifold.}
\end{figure}

As we observed in the case of $N=2$,
the gluing pattern can be regarded as the permutation of vertices
resulting from the permutation of the $S$ matrices (i.e. different indefinite
Toda lattices). To see this, we associate each component with an $S$ matrix
through
$sgn(\beta_i)=s_is_{i+1}$. For instance, $S=diag(1,-1,1)$ corresponds to $--$.
Then the gluing pattern gives for example that the marking of the component
associated with $S=diag(s_1,s_2,s_3)$ of the vertex $V[3,2,1]$
is the same as the one with $S=diag(s_2,s_1,s_3)$ of $V[2,3,1]$.
In general, let $V_1$ and $V_2$ be two vertices such that $PV_1P^{-1}
=V_2$, then $S=diag(s_1,s_2,s_3)$ of $V_1$ is marked the same as
$PSP^{-1}$ of $V_2$. In this sense, the symmetry group of the
compactified manifold is the {\it semi}-direct product of $S_3$ and $(\ZZ_2)^2$
\cite{dj}.
This manifold is to be compared with the $N=3$ Tomei's manifold, i.e.,
the iso-spectral manifolds of tridiagonal symmetric matrices $L_T$
in (\ref{l0}).
There are also four components giving hexagons
depending on the signs of $b_1$ and $b_2$.
The Toda
flow with $L_T$ is complete (no blow-up), and each component
is invariant under the flow. So no
gluing is neccessary, and the marking of the components is just given by the
 signs of $b_1$ and $b_2$ i.e. no permutation on the signs.
Thus the symmetry group for the case of Tomei is the {\it direct}
product of $S_3$ and $(\ZZ_2)^2$.

\medskip

With the above marking of the components, the flows in $B, C$ or $D$
generically blow up {\it twice} in
$t\in {\Bbb R}$.  This can be verified by
finding the zeroes of the $\tau$-functions $\tau_1$ and $\tau_2$:
Writing $\tau_1(t)=\sum_{i=1}^{3}\rho_ie^{\lambda_i t}$, $\tau_2(t)$ is given
by
\begin{eqnarray}
\nonumber
\tau_2(t)=\rho_1\rho_2(\lambda_1-\lambda_2)^2e^{(\lambda_1+\lambda_2)t}+
\rho_1\rho_3(\lambda_1-\lambda_3)^2e^{(\lambda_1+\lambda_3)t}+
\rho_2\rho_3(\lambda_2-\lambda_3)^2e^{(\lambda_2+\lambda_3)t}\ .\\
\label{tau2}
\end{eqnarray}
Without loss of generality we take $\rho_1>0$. Then there are four cases
depending on the signs
of $\rho_2$ and $\rho_3$:

\medskip

(1) $\rho_2>0$, $\rho_3>0$. In this case, $\tau_1(t)$ and $\tau_2(t)$
are positive definite. This corresponds to the flows in the component $A$.

(2) $\rho_2>0$, $\rho_3<0$. Since $\tau_1(t)>0$ for large postive $t$
and $\tau_1(t)<0$ for large negative $t$, there is at least one zero for
$\tau_1$. On the other hand, we find $(\tau_1e^{-\lambda_3t})'>0$,
so there is
only one simple zero for $\tau_1(t)$. Similarly, one shows that there is only
one simple zero for $\tau_2$. Since $\tau_2=\tau_1\tau_1''-(\tau_1')^2$,
these zeroes do not coincide.

(3) $\rho_2<0$, $\rho_3<0$. We show similarly as for (2) that both
$\tau_1$ and $\tau_2$ have a nondegenerate zero.

(4) $\rho_2<0$, $\rho_3>0$. This case has the most interesting feature as we
will see. From the asymptotic behaviors,
it is easy to see both $\tau_1$ and $\tau_2$
are convex in $t$ and have either no or two zeroes. To be more precise, we
consider $(\tau_1 e^{-\lambda_3 t})'=0$, and find the root,
\begin{eqnarray}
\label{t0}
t_0=\frac{1}{\lambda_1-\lambda_2}\ln \left[\frac{-(\lambda_2-\lambda_3)\rho_2}
{(\lambda_1-\lambda_3)\rho_1}\right]\ .
\end{eqnarray}
Then $\tau_1(t_0)$ becomes
\begin{eqnarray}
\label{tau1t0}
\tau_1(t_0)=\frac{\lambda_1-\lambda_2}{\lambda_2-\lambda_3}
\rho_2 e^{\lambda_2 t_0} + \rho_3e^{\lambda_3 t_0} \ .
\end{eqnarray}
Also considering $(\tau_2e^{-(\lambda_1+\lambda_2)t})'=0$, we find that
 the solution is exactly that given by (\ref{t0}), and
  $\tau_2(t_0)$ becomes
\begin{eqnarray}
\label{tau2t0}
\tau_2(t_0)=\rho_1e^{\lambda_1 t}(\lambda_1-\lambda_2)(\lambda_2-\lambda_3)
\tau_1(t_0) \ .
\end{eqnarray}
 From (\ref{tau1t0}) and (\ref{tau2t0}), if $\tau_1(t_0) \ge 0(<0)$,
then $\tau_2(t_0) \ge 0(<0)$. This implies that (i) if $\tau_1$ (or $\tau_2$)
has two simple zeroes then $\tau_2$ (or $\tau_1$) has no zero, or (ii)
both $\tau_1$ and $\tau_2$ have a doubly degenerated zero at $t_0$. Thus the
total number of zeroes in $\tau_1$ and $\tau_2$ functions is
generically given by two,
i.e. $1+1=2+0=0+2=1\times (3-1)$ (Proposition \ref{nzero}).

\medskip

For the general case, a direct analysis of $\tau$-function
 seems to be very hard for even counting the total number of zeroes in
 the $\tau$-functions. Instead, we study a flow passing through near the
 edges from the top vertex to the bottom vertex, and counts the number of
 blow-ups in the flow to determine the number of zeroes in $\tau$'s.

\medskip

The general case with $N\times N$ matrix $L_H$ can be obtained by a direct
extention of the previous examples. Here we summarize the result with
 some convenient notations.
There are $N!$ critical points of the flow giving the vertices of the
integral manifold of the Toda system, and we denote them as
\begin{eqnarray}
\label{vertex}
V[i_1,\cdots,i_N] := \left(
\begin{array}{lllll}
\lambda_{i_1} & 1  & 0 & \cdots & 0 \\
0 & \lambda_{i_2} & 1 & \cdots & 0 \\
\vdots & \ddots & \ddots & \ddots & \vdots \\
0 & \cdots & \ddots & \lambda_{i_{N-1}} & 1 \\
0 & \cdots & \cdots & 0 & \lambda_{i_N}\\
\end{array}
\right).
\end{eqnarray}
We denote the edges connecting the vertices $V[i_1,
\cdots,i_N]$ and $V[i_1,\cdots,i_{k+1},i_{k},
\cdots,i_N]$ by
\begin{eqnarray}
\label{edge}
E_k^{\pm}[i_1,\cdots,i_N]:= \left(
\begin{array}{lllll}
\lambda_{i_1} & 1  & 0 & \cdots & 0 \\
0 & \lambda_{i_2} & 1 & \cdots & 0 \\
\cdot & \ddots & \ddots & \ddots & \cdot \\
\vdots & \vdots & A[i_k,i_{k+1}] & \vdots & \vdots \\
\cdot & \cdot & \ddots & \ddots & \cdot \\
0 & \cdots & \cdots & \lambda_{i_{N-1}} & 1 \\
0 & \cdots & \cdots & 0 & \lambda_{i_N}\\
\end{array}
\right),
\end{eqnarray}
where $A$ is a $2\times 2$ matrix defined by
\begin{eqnarray}
\label{a22}
A[i_k,i_{k+1}]=\left(
\begin{array}{cc}
\alpha_{i_k} & 1 \\
\beta_{i_k} & \alpha_{i_{k+1}} \\
\end{array}
\right)
\end{eqnarray}
having $\lambda_{i_k}$ and $\lambda_{i_{k+1}}$ as the eigenvalues
and the superscript ${\pm}$ in the edge is assigned by $sgn(\beta_{i_k})$.
As we know from the case $N=2$, the flow on $E_k^{+}[i_1,\cdots,i_N]$
has no singularity,
while $E_k^{-}[i_1,\cdots,i_N]$ is glued at infinity with $E^{-}_k[i_1,
\cdots,i_{k+1},i_k,\cdots,i_N]$ denoted as $E_{{\underline k}}^-$ in Fig.4.
We call
$E_k^{-}[i_1,\cdots,i_N]$ an ``outgoing'' edge
if $\lambda_{i_k}<\lambda_{i_{k+1}}$, and
an ``incoming'' edge otherwise.
Around each vertex, there are
$2^{N-1}$ different components depending on the signs of $\beta_i$'s,
that is, these components are separated by the hypersurfaces ${\cal H}_k$
defined by $\beta_k=0$ for $k=1,\cdots,N-1$.
We assume that these hypersurfaces
intersect transversally in a neighborhood of each vertex.  In paticular,
the edges $E^{\pm}_k$ transversally intersect to the surface ${\cal H}_k$,
 and the superscript $\pm$
indicates the sides of the component separated by ${\cal H}_k$.
We denote these components
by $S[s_{1,2},\cdots,s_{N-1,N}]$ where $s_{k,k+1}:=s_ks_{k+1}$
(note then $sgn(\beta_k)=s_{k,k+1}$), and
also define
\begin{eqnarray}
\nonumber
S_k^{\pm}[s_{1,2},\cdots,s_{k-1,k},s_{k+1,k+2},\cdots,s_{N-1,N}]
:=S[s_{1,2},\cdots,s_{k-1,k},\pm,s_{k+1,k+2},\cdots,s_{N-1,N}],\\
\label{spm}
\end{eqnarray}
which are the sections of the components $S$ devided by the
surface ${\cal H}_k$, and $S=S^+_k\cup S^-_k$.
Since the edge $E^+_k[i_1,\cdots,i_N]$ connects the vertices $V[i_1,\cdots,
i_{k},i_{k+1},\cdots,i_N]$ and $V[i_1,\cdots,i_{k+1},i_k,\cdots,i_N]$ without
a singularity, the sections $S_k^+$ of these vertices coincide.  However
the sections $S^-_k$ of these vertices are not connected identically along with
the signature of $[s_{1,2},\cdots,s_{N-1,N}]$, but with permutation on
the signs in the $S$ matrices as we showed in the case of $3\times 3$ matrix.

\medskip

We now show that the connection of these sections through blow-up is
completely determined by tracing the flows of the Toda lattices with $L_H$
(i.e. this may be referred as a compactification of the iso-spectral
manifold of tridiagonal Hessenberg matrices by the orbit gluing).
More precisely, let a flow of $L_H$ starts in some component of
a vertex, then after blow-up, we need to determine which component
of other vertex the flow goes in.
The key of our approach is to use continuity of $D_i(t)$'s (or $\tau_i(t)$'s)
on the initial data,
which can be easily seen from (\ref{DDD}) and (\ref{tau}).
Also from relations (\ref{atau}) and (\ref{btau}), the flow $L_H(t)$ is
continuous on the intial data except the times of blow-ups given by the
zeroes of $D_i(t)$'s (or $\tau_i(t)$'s).
Then we have:
\begin{Lemma}
\label{continuity}
Suppose $t_0, \cdots, t_m$ are all zeroes of $D_i(t)$'s ($\tau_i(t)$'s)
with initial data $L_H^0$. Then for any finite $\tilde t$ with
$\tilde t \neq t_k$ for $k=0,\cdots,m$, there exists a neighborhood
of $L_H^0$, such that its time evolution at $\tilde t$ is a neighborhood of
$L_H(\tilde t)$.
\end{Lemma}

We now study the flows around the edges. If the superscript of an edge
denoted by (\ref{edge}) is ``+'', then there is no blow up in the flow.
Suppose an initial data $L_H^0$ is on $E_k^-[i_1,\cdots,i_N]$.
In a small neighborhood of $L_H^0$,
we can take a hypersurface ${\cal G}_k$ of $L_H^0$ transversal to the flow
with codimension one. Since the edge $E_k^-[i_1,\cdots,i_N]$ is shared
by all the components $S^-_k[s_{1,2},\cdots,s_{k-1,k},s_{k+1,k+2},
\cdots,s_{N-1,N}]$, ${\cal G}_k$ can be devided into
these $2^{N-2}$ sections.
If $\lambda_{i_k}<\lambda_{i_{k+1}}$, $L_H(t)$ blows up at some $t_0$,
then it jumps to edge $E_k^-[i_1,\cdots,i_{k+1},i_k,\cdots,i_N]
\equiv E_{{\underline k}}^-$ as indicated in Fig.4.
By Lemma \ref{continuity},
for some $t>t_0$, we take ${\cal G}_k$ sufficiently small such that
${\cal G}_k$ evolves to
a hypersurface ${\cal G}_k'$ at $L_H(t)$. We are interested in how
sections of ${\cal G}_k$
are glued with those of ${\cal G}_k'$. We have the following proposition:

\begin{Proposition}
\label{main}
Every point of the section $S^-_k[s_{1,2},\cdots,s_{k-1,k},s_{k+1,k+2},
\cdots,s_{N-1,N}]$ on the hypersurface ${\cal G}_k$ is glued
by the Toda flow with a unique point in the section marked by
$S^-_k[s_{1,2},\cdots,
s_{k-1,k+1},s_{k,k+2},\cdots,s_{N-1,N}]$($\equiv S_{{\underline k}}^-$
in Fig.4) after the blow-up, that is,
the gluing pattern is
given by the permutation of the signs $s_k \leftrightarrow s_{k+1}$.
\end{Proposition}
\begin{Proof}
We calculate directly $D_i(t)$'s with
the initial data $L_H^0$ given on the edge $E_k^-$ as
\begin{eqnarray}
\nonumber
D_j(t)&=&\exp (\sum_{l=1}^j\lambda_{i_l}t), \quad {\hbox {for\ }} j\neq k \\
\label{dedge}
D_k(t)&=&(-\sinh^2\mu_0 \ e^{\lambda_{i_k}t}+\cosh^2\mu_0 \
e^{\lambda_{i_{k+1}}t}) \exp (\sum_{l=1}^{k-1}\lambda_{i_l}t) \ .
\end{eqnarray}
 From (\ref{dedge}), we see $D_j(t)$ are positive definite for $j\neq k$,
while $D_k(t)$ has one zero at some $t_0$.
By continuity of $D_i(t)$'s, ${\cal G}_k$ can be chosen such that all the
flows starting
from ${\cal G}_k$ have $D_k(t)$ reaches zero first and the zero is
nondegenerate.
So from (\ref{taud}), $\tau_k(t)$ reaches zero first. From (\ref{btau}),
the sign change of $\tau_k(t)$ implies the sign changes of $\beta_{k-1}$
and $\beta_{k+1}$, that is,
$S^-_k[s_{1,2},\cdots,
s_{k-1,k},s_{k+1,k+2},\cdots,s_{N-1,N}]$ connects to
$S^-_k[s_{1,2},\cdots,
s_{k-1,k+1},s_{k,k+2},\cdots,s_{N-1,N}] \equiv S_{{\underline k}}^-$, as
 shown in Fig. 4.
\end{Proof}


\begin{figure}
\epsfysize=7cm
\centerline{\epsffile{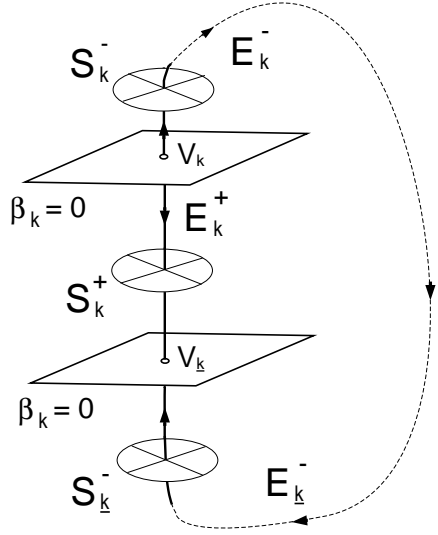}}
\caption{The gluing pattern of the component $S_k$ through the flow.}
\end{figure}

This proposition gives a clear picture of the gluing of orbits around edges.
In some sense, components near an outgoing edge stay close to the edge
and connect to
appropriate components near the incoming edge.
For an orbit whose first zero of $\tau_i(t)'s$
is nondegenerate, by Lemma \ref{continuity},
there exists a neighborhood which will blow up and evolve to the
same component as the orbit.
So to see the gluing pattern away from an edge $E^-$, we do continuation
starting from the edge. This process stops when the first zero of some orbits
becomes degenerate.
In this way, we obtain a maximum collection of orbits
which follow $E^-$. We denote this set by $C_E$.
In general, a component $S[s_{1,2},\cdots,s_{N-1,N}]$ may have several outgoing
edges $E^-_1,\cdots,E^-_r$. We do continuation for each of them. Then we claim
that generic orbits follow some edge in the following sense:

\begin{Proposition}
\label{continuation}
Suppose $S$ is a component with nonzero outgoing edges $E^-_1,\cdots,E^-_r$.
$C_{E_1},\cdots,C_{E_r}$ as defined above. Then
$S=\overline {\bigcup _{k=1}^rC_{E_k}}$, i.e.,
the closure of the union of $C_{E_k}$'s.
\end{Proposition}
\begin{Proof}
We prove by contradiction. Suppose the proposition is not true, that is,
 there is some region $R$ in $S$
separated from edges, i.e., the boundary of $R$ has a degenerate first zero.
By going forward in time, $R$ blows up to some region of some component.
This region must also be separated from edges, otherwise, we can apply
Lemma \ref{continuity} around the edge back in time,
which contradicts $R$ separated from edges. Eventually, $R$ reaches
some component of the bottom vertex $V[1,\cdots,N]$, and
no more blowup after that.
Suppose $\tilde R$ is the second last region $R$ visits. Since the boundary
of $\tilde R$ has degenerate zeroes, it will change to a different component
from $\tilde R$ does. But by Lemma \ref{continuity}, for large $t$,
the boundary
should get together with $\tilde R$, which is a contradiction.
\end{Proof}

\medskip

With Propositions \ref{main} and \ref{continuation}, we have a complete
pattern of gluing which provides the compactified manifold of the
solution of indefinite Toda latticies.
{From} the gluing pattern, we now prove Proposition \ref{nzero}.

\medskip

{\it Proof of Proposition \ref{nzero}}.
First we determine the signs of $\beta_i(t)$'s for $t \rightarrow
-\infty$ by studying asymptotic behaviors of $\tau_i(t)$'s. From
(\ref{btau}) and (\ref{expand}), we have for large negative $t$
\begin{eqnarray}
\nonumber
sgn(\beta_i(t))&=& sgn\left(\frac{(\rho_N\cdots \rho_{N-i})(
\rho_N\cdots \rho_{N-i-2})}{(\rho_N\cdots \rho_{N-i-1})^2}\right) \\
\label{signb}
&=&sgn(\rho_{N-i-1}\rho_{N-i-2})=s_is_{i+1} \ .
\end{eqnarray}
So $s_i$ can be taken as $s_i=sgn(\rho_{N-i-2})$, thereby the number
of $s_j=-1$ is also $m$. Now we start in component
$S[s_{1,2},\cdots,s_{N-1,N}]$ of
the top vertex $V[N,\cdots,1]$, since generic orbits follow edges, all we
need to do is to count how many blow-ups are needed through the edges from
$V[N,\cdots,1]$ to $V[1,\cdots,N]$.
We count by induction on $N$. For $N=2$, the Proposition holds.
Suppose it is valid for $N-1$.  If $s_N=1$, it takes  $m$ blow-ups
to get to the top, then the system is reduced to $N-1$. The total number of
zeroes is thus $(N-1-m)m +m=m(N-m)$. The case $s_N=-1$ can be shown
similarly.

\medskip


\section{CW decomposition and nonorientability}
\renewcommand{\theequation}{5.\arabic{equation}}\setcounter{equation}{0}

In this section, we give a CW decomposition of the compactified manifold
$M_N$ obtained in the previous section, and show that $M_N$ is
not orientable for $N>2$.
For the iso-spectral manifolds in the case
of the tridiagonal symmetric matrices, a CW decomposition was given
in \cite{tomei}.
In our case,
the $j$-cells are given by the glued components marked by the signs of
$\beta_i$'s with $j$ signs and  $N-j-1$ zeroes, say,
$\beta_{k_1}=\beta_{k_2}=\cdots=\beta_{k_{N-j-1}}=0$.
Let $S_V^j$ be a component associated with a vertex V of a $j$-cell.
Then with the gluing pattern discussed in the previous section,
the $j$-cell $\hat S^j$ can be written as
\begin{eqnarray}
\label{jcell}
\hat S^j=\bigcup_P
(PS_V^jP^{-1})_{PVP^{-1}}\bigcup \{\infty\}\ ,
\end{eqnarray}
where $P\in S_N$ keeping
$\lambda_{k_1}, \cdots, \lambda_{k_{N-j-1}}$ fixed.
For example, in the case of $N=3$,
the vertices are 0-cells, the edges are 1-cells, and
the hexagons are 2-cells. Then the hexagon marked by $B$ in Fig. 3 is
expressed by a union of
$S[1,-1]_{V[3,2,1]}$, $S[-1,-1]_{V[3,1,2]}$,
$S[-1,-1]_{V[2,1,3]}$ and $S[-1,1]_{V[1,2,3]}$ together with infinities.
In particular, we denote
$\hat S[s_{1,2}, \cdots, s_{N-1,N}]$ as the ($N-1$)-cell with a
component
$S[s_{1,2}, \cdots, s_{N-1,N}]$
around the top vertex $V[N,\cdots,1]$.
For example, the 2-cell $B$ above is denoted by $\hat S[1,-1]$.
Then we have:

\begin{Proposition}
\label{homology}
$M_N$ is not orientable, i.e.
$H_{N-1}(M_N, Z)=0$, for $N > 2$.
\end{Proposition}
\begin{Proof}
There are $2^{N-1}$ ($N-1$)-cells. First we look at them around $V[N,\cdots,
1]$.
Since two components differ by one sign have a common ($N-2$)-face
around $V[N,\cdots,1]$, in order to cancel the boundary, they must have
opposite orientations. So starting with $S[1,1,\cdots,1]$, the only
combination of cells possible to have zero boundary is
\begin{eqnarray}
\label{c}
\hat C=\sum_{(s_1,\cdots,s_N)}(-1)^{n(s_1,\cdots,s_N)}\hat
S[s_{1,2},\cdots,s_{N-1,N}]\ ,
\end{eqnarray}
where $n(s_1,\cdots,s_N)$ is the number of minus signs in
$(s_{1,2},\cdots, s_{N-1,N})$. However, any component around $V[N,\cdots,1]$
is carried by flow to become adjacent to $S[1,1,\cdots,1]$ around
some vertices. To see this, note that $(s_1,\cdots,s_n)$ can be permutated
to be (1,$\cdots$,1,$-1,\cdots,-1)$.
Since in (\ref{c}) there are cells having same orientation as
$S$[1,1,$\cdots$,1] for $N>2$, it is impossible for $\hat C$ to have zero
boundary.
\end{Proof}

\par\medskip\medskip

{\bf Acknowledgment}
The authors would like to thank M. Davis and T. Januskiewicz
for valuable discussions, especially, for pointing out the symmetry group
of the manifolds. The authors would also like to thank B. Baishanski
for suggesting the reference \cite{karlin}, and B. Luce for
making Figures.
The work of Y.K. is partially supported by an NSF grant DMS9403597.


\bibliographystyle{amsplain}

\end{document}